\documentclass[journal=jctc,manuscript=article]{achemso}

\usepackage{chemformula} 
\usepackage[T1]{fontenc} 
\usepackage[utf8]{inputenc}
\usepackage{bbm}
\usepackage{nameref}

\usepackage{color}

\author{Verena Kristin Gupta}
\affiliation{Bremen Center for Computational Materials Science, University of Bremen, P.O. Box 330440, D-28334 Bremen, Germany}
\author{B\'alint Aradi}
\affiliation{Bremen Center for Computational Materials Science, University of Bremen, P.O. Box 330440, D-28334 Bremen, Germany}
\author{Kyoung Kweon}
\affiliation[llnl]{Physical and Life Sciences Directorate, Lawrence Livermore National Laboratory, Livermore, California 94550, United States}
\author{Nathan Keilbart}
\affiliation[llnl]{Physical and Life Sciences Directorate, Lawrence Livermore National Laboratory, Livermore, California 94550, United States}
\author{Nir Goldman}
  \affiliation[llnl]{Physical and Life Sciences Directorate, Lawrence Livermore National Laboratory, Livermore, California 94550, United States}
  \alsoaffiliation{Department of Chemical Engineering, University of California, Davis, California 95616, United States}
\author{Thomas Frauenheim}
\affiliation{Computational Science Research Center, No.10 East Xibeiwang Road, Beijing 100193 and Computational Science and Applied Research Institute Shenzhen, China.}
\alsoaffiliation{Bremen Center for Computational Materials Science, University of Bremen, P.O. Box 330440, D-28334 Bremen, Germany}
\email{thomas.frauenheim@bccms.uni-bremen.de}
\author{Jolla Kullgren}
\affiliation{Department of Chemistry, Structural Chemistry, Angstr\"om Laboratory, Uppsala University, Box 538, 752 21, Uppsala, Sweden}
\email{jolla.kullgren@kemi.uu.se}

\title[DFTB nanocrystal]
  {Using DFTB to Model Photocatalytic Anatase-Rutile TiO$_2$ Nanocrystalline Interfaces and their Band Alignment}

\abbreviations{DFT,DFTB}

\begin{document}







\begin{abstract}
Band alignment effects of anatase and rutile nanocrystals in TiO$_2$ powders lead to an electron hole separation, increasing the photo catalytic efficiency of these powders. While size effects and types of possible alignments have been extensively studied, the effect of interface geometries of bonded nanocrystal structures on the alignment is poorly understood. In order to allow conclusive studies of a vast variety of bonded systems in different orientations, we have developed a new density functional tight binding parameter set to properly describe quantum confinement in nanocrystals. Applying this set we found a quantitative influence of the interface structure on the band alignment.
\end{abstract}

\section{Introduction}
TiO$_2$ powders commonly used for photo catalytic applications such as the Degussa P25 (DP25) consist of anatase and rutile nanocrystals in different compositions (commonly $\sim 80$\% anatase and $\sim 20$\% rutile). This composition of the two different TiO$_2$ phases is shown to have a great impact on the photo catalytic efficiency due to band alignment effects that lead to an effective electron-hole pair separation. There have been several theoretical attempts to predict the type of band alignment and thus which crystal acts as a hole or electron trap\cite{PeterBA,slabBA,nanomain,nano2}. Experiments indicate that for the case of DP25 anatase acts as an electron trap, while rutile as a hole trap\cite{baexp}. Theoretical works have so far been focused on isolated bulk\cite{PeterBA}, slab \cite{slabBA} or nanocrystal\cite{nanomain,nano2} systems. While bulk studies predict a trapping of electrons on anatase and a trapping of the holes on rutile in accordance with the experiment, nanocrystal studies predict a significant size dependence for the alignment until bulk-like behavior is reached. However, neither of those approaches consider the potential impact of different types of additional degrees of freedom in materials such as geometrical interface arrangements. 
The role of such interface arrangements were investigated in Ref.~\citenum{slabBA} using slab models. However, such models can only give limited insights in the interface effects as the slabs of the two phases must be commensurate, allowing only for a few possible orientations and surface geometries. This makes it impossible to explore the degrees of freedom in the interface formation, which are present under experimental conditions at the microscale.

The vast phase space of possible nanocrystal orientations and sizes thus makes computational studies of anatase-rutile nanocrystal interfaces very challenging, even for very small nanocrystal models where a conclusive study on the role of the interfaces would be unfeasible using DFT based calculations alone. Consequently, we have developed a density functional tight binding (DFTB) parameter set to efficiently simulate TiO$_2$ nanocrystals and anatase-rutile interfaces. DFTB, being 2-3 orders of magnitude faster than comparable DFT calculations, provides the possibility of studying a vast number of systems and system sizes. Additionally, with the right choice of parameters, DFTB retains most of the accuracy of DFT and thus yields the possibility of making a closer to one-to-one comparison to experiments. Another benefit of the DFTB approach is its localized basis set which simplifies the computation of non-periodic structures such as nanocrystals.

In the present work we identify several issues in describing nanocrystals with the previously published DFTB parameter sets for titanium dioxide \cite{matsci1,tiorg}, e.g.\ problems in describing the quantum confinement effects correctly and yielding wrong relaxations when under-coordinated species are involved. The latter problem has been also reported in Reference \citenum{Balzaretti2021}. To remedy those issues, we have developed a new parameter set, named \textit{tio2nano} \cite{tio2nano}, that focuses on the correct description of TiO$_2$ nanocrystals and interfaces. It is based on the 3ob-parameter set \cite{3ob-gaus}. The electronic parameters of the new Ti species were tailored to ensure the proper description of the quantum confinement effects in TiO$_2$ nanocrystals. Instead of the traditional two-center potentials, we have used a many-body force field \cite{Koziol17,Lindsey17} to represent the repulsive contributions to the total energy, following the same approach as in Ref.~\citenum{Goldman_DFTB_H_Pu}. The three-center contributions of the force field allowed us to obtain improved agreement with DFT optimized geometries, compared to results from two-center repulsive potentials only.

For the band alignment studies, we considered explicit anatase-rutile interface structures. We are not aware of any previous attempts to determine the electronic structure using such models, but similar CeO$_2$ nanoparticle models were addressed in Ref. \citenum{fchem}.
By applying the newly developed DFTB parameter set on the generated interface structures, we were able to demonstrate the impact of the geometric alignment of the crystals on the band alignment.

\section{Methods}\label{sec:method}

DFTB calculations were performed with the DFTB+ code \cite{dftb+}. As described below in detail, the 3ob parameter set \cite{3ob-gaus} has been extended with the element Ti by using a density compression radius of 8.5~Bohr and wave compression radii of 5.609, 3.958 and 7.0~Bohr for the s-, p- and d- orbitals of the Ti-atom, respectively. The energy and the chemical hardness (a.k.a.\ the Hubbard-U value) of the virtual 4p orbital of the Ti-atom have been set to 0.206~Ha and -0.08~Ha, respectively.

The energies were created using the Chebyshev Interaction Model for Efficient Simulation (ChIMES)\cite{Koziol17,Lindsey17}, a reactive many-body molecular dynamics (MD) force field. ChIMES creates many-body interactions by projecting DFT computed data (e.g., forces, stress tensors, energies) onto linear combinations of many-body Chebyshev polynomials\cite{Pham_HN3_2021}. Briefly, this begins with an N-body expansion of the total energy for a system:
\begin{equation}
\label{echimes}
E_{\mathrm{rep}} = E_{\mathrm{ChIMES}}=\sum_i^N E_i + {\sum_i^N}\sum_{j>i}^NE_{ij}+\sum_i^N\sum_{j>i}^N\sum_{k>j}^NE_{ijk} + \mathcal{O}(n).
\end{equation}
Here, $E_i$ is the single atom energy for a given element, $E_{ij}$ and $E_{ijk}$ represent the 2--body and 3--body interaction energy, respectively, $N$ is the total number of atoms in the system, and $\mathcal{O}(n)$ corresponds to higher order terms. 

Specifically, the 2--body interactions are expressed as a linear combination of Chebyshev polynomials of the first kind:
\begin{equation}
\label{2bodyE}
E_{ij} = f_{\mathrm{P}}^{ij}(r_{ij})+f_\mathrm{C}^{ij}(r_{ij})\sum_{n=1}^{\Theta_2}\mathbbm{c}^{e_ie_j}_{n}T_n(s^{e_ie_j}_{ij})
\end{equation}
In this case, $T_n(s_{ij})$ is the Chebyshev polynomial of the first kind of $n^{\mathrm{th}}$ order, $e_i$ and $e_j$ are the element types of atoms $i$ and $j$, and $s_{ij}$ is a transformation of the inter-atomic distance $r_{ij}$ over the Chebyshev interval of [-1,1] using a Morse-like function. In addition, $\Theta_2$ corresponds to the two-body polynomial order, $f_\mathrm{C}^{ij}(r_{ij})$ is the cut-off function that ensures the potential and its derivative vary smoothly to zero beyond a specified distance, and $f_\mathrm{P}^{ij}(r_{ij})$ is a penalty function~\cite{Goldman_DFTB_H_Pu,LCO-2020} that helps to prevent sampling of inter-atomic distances below those seen in the training set (see Ref.~\citenum{Lindsey_CO_2020} for further details). The $\mathbbm{c}^{e_ie_j}_n$ are the set of permutationally invariant coefficients of linear combination for a given atom pair-type that are determined via a linear least-squares method.  

Similarly, 3--body interaction energies are expressed as a product of Chebyshev polynomials for the constituent atom pairs of a given triplet, yielding an orthogonal 3--body polynomial with $\binom{3}{2} = 3$ total interactions for a given triplet:
\begin{equation}
\label{3bodyE}
E_{ijk} = f_\mathrm{C}^{ij}(r_{ij}) f_\mathrm{C}^{ij}(r_{ik}) f_\mathrm{C}^{ij}(r_{jk})\sum_{m=0}^{\Theta_3} \sum_{p=0}^{\Theta_3}\sum_{q=0}^{\Theta_3}{'} \mathbbm{c}^{e_ie_je_k}_{mpq} T_m(s^{e_ie_j}_{ij}) T_p(s^{e_ie_k}_{ik}) T^{e_je_k}_q(s^{e_je_k}_{jk}). 
\end{equation}
In this case, $\Theta_3$ is used to label the three-body polynomial order with a single permutationally invariant coefficient $\mathbbm{c}^{e_ie_je_k}_{mpq}$ for each set of triplet atom types. To guarantee that only 3-body interactions between $i, j, k$ are counted towards the sum, only terms for which at least two of the three $m,p,q$ indices are greater than zero are included in the sum (indicated by the prime in Eq.~\ref{3bodyE}). In this way, many-body interactions can be included in the repulsive energy ($E_\mathrm{rep}$) by solving a linear least-squares optimization problem where optimal coefficients of the linear combination are determined directly. This avoids reliance on iterative approaches that are required for non-linear optimization problems (e.g., Levenberg-Marquardt) that are usually more computationally time consuming and not guaranteed to result in the global minimum. ChIMES models have been extended to include 4--body interactions in a similar fashion in reactive MD simulations\cite{Pham_HN3_2021,Lindsey_AL_2020}, though truncation of the ChIMES total energy with the 3--body term has proven sufficient for determination of the DFTB repulsive energy. 

Our training set for the ChIMES force field was determined from DFT-MD simulations of amorphous TiO$_2$ run at temperatures of 2250 and 300 K using the VASP code \cite{VASP}. For these calculations, we used the PBEsol functional\cite{PBEsol} with an energy cutoff of 550~eV and $\Gamma$-point sampling of the Brillouin zone. PBEsol was chosen due to its improved description of solids and their surfaces. The amorphous phase allowed for improved sampling of a wide-range of interatomic distances over the short-time scales of the simulations, which was enhanced by including data from elevated temperature. Each MD simulation was run on a system of 216 atoms for a total of 5~ps, with configurations taken for $E_\mathrm{rep}$ training every 100~fs (in order to allow for decoupling between training data), resulting in 50 training configurations taken from each temperature. We also included additional 10 configurations from an MD simulation run at 40 GPa in order to improve sampling of close interatomic distances. This yielded a total of 110 training configurations. 

The training set to determine $E_\mathrm{rep}$  was then computed by subtracting the gradient of the DFTB electronic energy from our DFT reference property, i.e., $\vec{F}_\mathrm{TRAIN} = \vec{F}_{DFT}-\vec{F}_\mathrm{DFTB-elec}$. In this work, a spline fit of Ti-Ti dimer repulsive interactions was included in the DFTB electronic calculations order to ensure that excessively small inter-atomic distances were not approached during calculations. In addition, O-O distances were poorly sampled in our training set and these repulsive parameters were thus taken from the 3ob-0-1 parameter set and were not a part of our fit. ChIMES parameters were determined through linear least-squares fitting to the resulting ionic forces and the diagonal components of the stress tensor for each of these MD snapshots. We set the ChIMES 2--body polynomial order to 12 and the 3--body order to 8, similar to previous work\cite{Dantana_OPV_2020}, and solve for optimal coefficients using the Least-Angle Regression (LARS)\cite{lars1,lars2} algorithm with a Least Absolute Shrinkage and Selection Operator (LASSO)\cite{lasso} regularization value of $10^{-4}$. 

For the DFTB calculations, charges were converged with a tolerance of $10^{-6}$ a.u. and forces with $10^{-4}$ a.u. Bulk calculations were performed with an 8$\times$8$\times$8 Monkhorst-Pack \cite{MonkhorstPack} grid for the primitive unit-cell. The rutile (110) surface was constructed as a 4$\times$2 surface unit cell with 6 layers, the (100) surface as a 4$\times$2 unit cell with 6 layers and the (001) surface as $4\times4$ with 8 layers. The (001) anatase surface had a 2$\times$2 unit cell with 8 layers, the (101) surface had a 8$\times$4 unit cell with 5 layers. All rutile structures were computed with a 2$\times$2$\times$1 k-point mesh, the anatase (001) surface with a 4$\times$4$\times$1 mesh and the (101) surface with $\Gamma$-only calculations. Branching point energy calculations (discussed below) were performed with the primitive unit cell of the anatase and rutile crystals using an 8$\times$8$\times$8 k-point grid. For nanocrystals we performed $\Gamma$-only calculations.

The nanocrystal interfaces were constructed using the JANUS code , which is based on a quick hull algorithm for convex hulls \cite{janus}. This algorithm allows to rapidly identify convex hulls from a set of points, in this case the coordinates of atoms from the nanocrystals in xyz format. The code then combines simplices of the computed convex hull for all normals that form an $\arccos$ of $0.99$ or larger. The largest facets of each individual nanocrystal are then aligned to each other so that their surface normals are parallel to each other and to the x-axis with a user specified distance between them. The procedure is repeated with the nanocrystals being rotated with respect to each other around the x-axis for angles between 0 and 180 degrees in steps of 15 degrees. Additionally, the nanocrystal gets displaced in the yz-plane along the perimeter of a circle with a user specified radius. This displacement occurs with angles between 0 and 360 degrees in steps of 30 degrees. This results in 156 interface geometries which are stored in an ASE \cite{ASE} trajectory format. To create the interfaces we used the two smallest relaxed nanocrystals of anatase and rutile. The nanocrystals were aligned with a distance along x of 2~\AA{} and a radius of 2~\AA{} was used for the displacement in the yz-plane.

Interface energies were computed as
\begin{equation}
    E_\mathrm{inter} = E_\mathrm{tot} - \left( E_\mathrm{rut} + E_\mathrm{ana}\right)
    \label{eq:einter}
\end{equation}
where $E_\mathrm{tot}$ is the energy of the interface structure, $E_\mathrm{rut}$ the energy of the isolated rutile nanocrystal contained in the interface structure and $E_\mathrm{ana}$ accordingly the energy of the anatase nanocrystal.

For comparative DFT calculations we used the Vienna Ab initio Simulation Package \cite{VASP} (VASP 5.4.4) with the PAW method to describe the cores. For titanium the 3p electrons were treated as valence electrons. The plane wave cut-off was chosen to be 420 eV, the augmentation cut-off was chosen as 840 eV. Geometry relaxations were performed until energy differences were converged below an error of $10^{-3}$ eV. The k-point meshes were chosen in accordance with the DFTB calculations.

As nanocrystal models we chose Wulff-type crystal structures for anatase nanocrystals which expose (101) facets and quasi Wulff-type crystals for rutile exposing (110), (100) and (101) facets. These nanocrystal structures were chosen in accordance with Ref.~\citenum{nanomain} to allow for direct comparison. Additionally, for reasons described below, we also considered another set of quasi Wulff-type rutile nanocrystals by cutting particles along the (110) and (101) planes with an even number of (110) layers across the waist of the particle. 

\section{Results and discussion}\label{sec:resdisc}

\subsection{Parameterization and validation of the new DFTB parameter set}\label{sec:dftb-param}

\begin{figure}[htbp]
    \centering
    \includegraphics[width=0.8\textwidth]{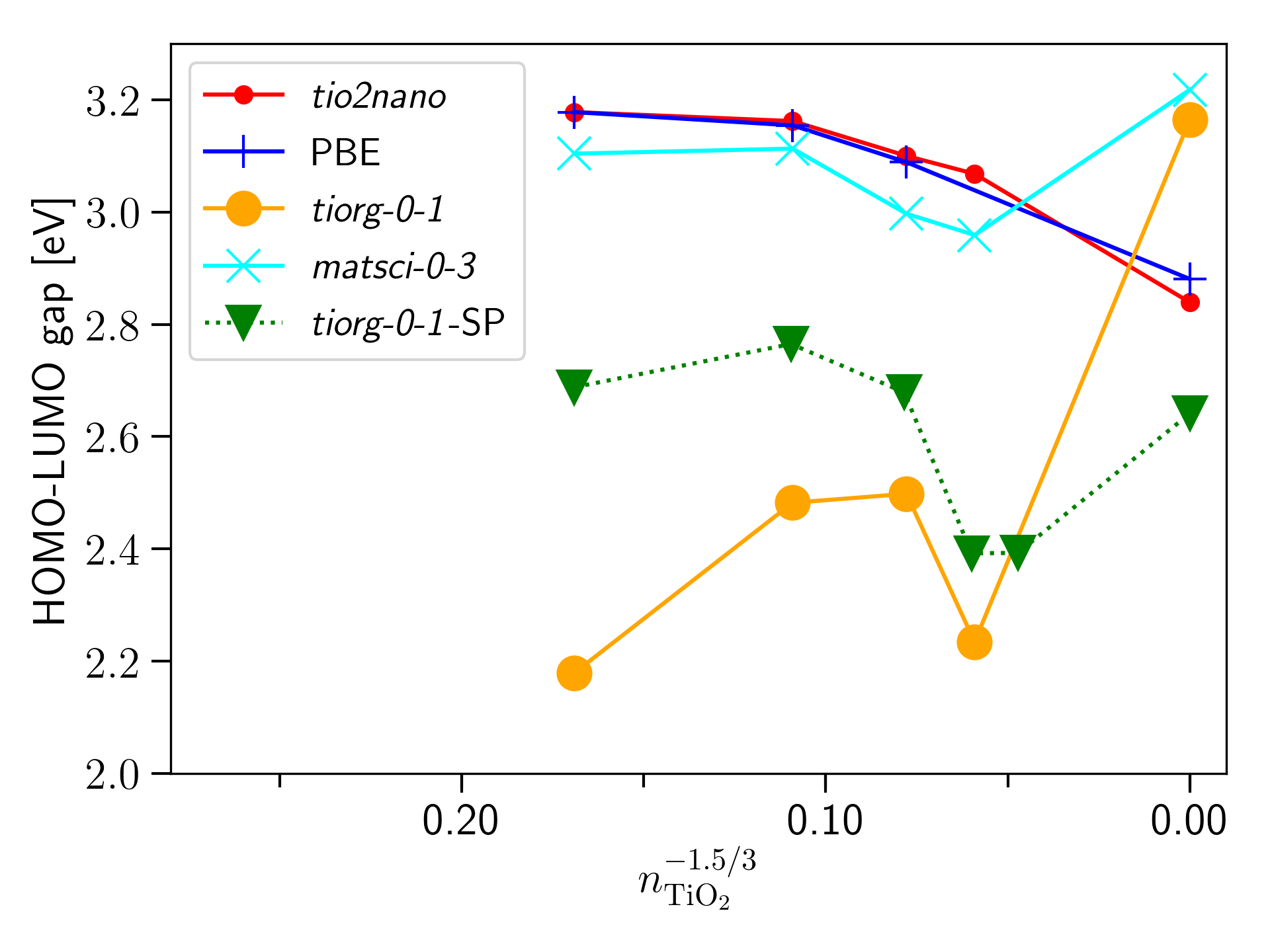}
    \caption{Calculated HOMO-LUMO gaps of anatase nanocrystals using PBE and DFTB with the \textit{tiorg-0-1}, \textit{matsci-0-3} and the newly developed \textit{tio2nano} set. The PBE results have been shifted to align with the values obtained by the new parameter set for the smallest nanocrystal. The \textit{tiorg-SP} values were obtained by carrying out DFTB calculations with the \textit{tiorg-0-1} set at the PBE geometries (without relaxing with DFTB).}
    \label{fig:dip-gap}
\end{figure}

The existing parameter sets for Ti-O interactions, the \textit{tiorg-0-1} \cite{tiorg} and the \textit{matsci-0-3} \cite{matsci1} sets, were evaluated only for bulk structures and common surfaces so far. By applying them to describe anatase Wulff-type nanocrystals, we faced the issue that the HOMO-LUMO gap did not change with the particle size as predicted by the theory of quantum confinement\cite{nanomain} and as observed in analogous PBE calculations. Instead of the expected linear function of $n^{-2/3}$ (with $n$ being the number of TiO$_2$ units) we observed surface states, which fall into the band gap and give rise to HOMO-LUMO gap sizes lower than the bulk value as shown in Fig.~\ref{fig:dip-gap}. In order to rule out geometrical effects, we also performed single point calculations for those nanocrystals using their PBE geometries, but found a similar wrong behavior with both sets. Efforts to resolve the issue by changing the compression radii of the Ti-atom turned out to be unsuccessful.
As a next attempt, we tried to describe the Ti-O interaction by extending the 3ob-parameter set \cite{3ob-gaus} with Ti. The electronic parameters of the Ti atom were optimized by fitting on the theoretically expected linear behaviour of the band gap with respect to $n^{-2/3}$. For this fit, anatase nanocrystals with sizes of $n=84$, $165$ and $286$ and the bulk phase of anatase had been used resulting in the compression radii reported in the \nameref{sec:method} section.
\begin{figure}[htbp]
    \centering
    \includegraphics[width=0.8\textwidth]{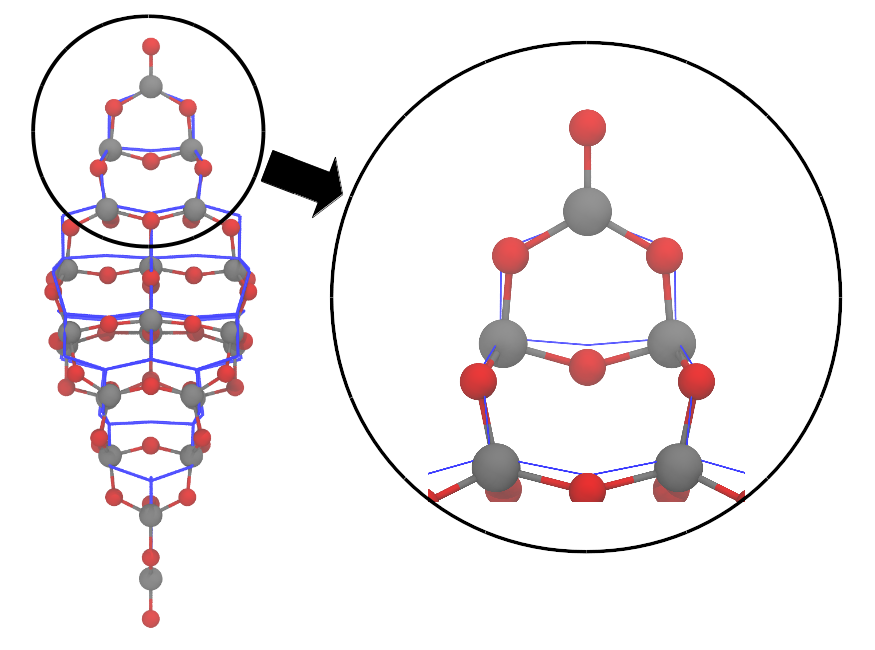}
    \caption{Relaxed geometries of the Ti$_{35}$O$_{70}$ nanocrystal. The DFTB geometry obtained using a two-body repulsive potential for the Ti-O interaction is shown by balls and sticks, while the PBE geometry is represented by blue solid lines. The geometries had been aligned at the top oxygen atom.
    }
    \label{fig:wrongrelax}
\end{figure}


After fixing the electronic parameters, the two-center repulsive potentials of the Ti-Ti and the Ti-O interactions had been derived using a symmetrized titanum hcp crystal and a symmetrized rutile bulk crystal as reference, respectively. The symmetrization ensures that the first neighbour Ti-Ti (in hcp Ti) and Ti-O bonds (in rutile) have identical lengths, allowing for a simple manual fitting procedure. When relaxing the nanocrystals with the obtained repulsive potentials, the resulting geometries exhibited significant deviations in the strongly undercoordinated tip regions as compared to PBE-geometries (see Fig.~\ref{fig:wrongrelax} for an example). Unfortunately, those geometrical changes lead to the appearance of spurious surface states in the gap. Our attempts to enforce the correct geometry by tuning the two-center repulsive functions were not successful. This is not surprising as the incorrect Ti-O-Ti angle seems to be the main driving force behind the differences in the relaxed geometries. To enforce correct nanocrystal geometries (and with that also the correct electronic structure), we decided to represent the Ti-O and Ti-Ti repulsive functions with the ChIMES force field, which allows the inclusion of three center terms. The interface to the ChIMES software had been implemented in a development version of the DFTB+ package. For the Ti-Ti repulsive potential we kept the two-center one derived from the symmetrized titanium hcp crystal, with ChIMES corrective terms. It is important to note, that the performance of the Ti-Ti repulsive potential in describing Ti bulk phases was not evaluated in this work.

The resulting parameterization performs well for anatase and rutile bulk as well as for their surfaces and is able to describe systems for which the previous parameter sets have failed. The bulk lattice parameters given in Table \ref{tab:bulk} show a very good agreement with ab initio counterparts for rutile. For anatase we observe a slight overestimation of the $a$ value, while $c$ is underestimated.
\begin{table}[h]
    \centering
    \begin{tabular}{c c c c c c}
        \hline
        & DFTB$^a$ & PBE$^a$ & PBEsol$^a$ & BLYP$^b$ & Exp.$^c$ \\ \hline
        rutile & & & & & \\
        $a$ & 4.629 & 4.621 & 4.600 & 4.679 & 4.594 \\
        $c$ & 2.980 & 2.954 & 2.940 & 2.985 & 2.959\\
        & & & & & \\
        anatase & & & & & \\
        $a$ & 3.887 & 3.792 & 3.780 & 3.828 & 3.784\\
        $c$ & 9.293 & 9.640 & 9.580 & 9.781 & 9.515\\
        \hline
    \end{tabular}
    \caption{Calculated and experimental anatase and rutile bulk lattice parameters (in \AA). The superscripts $a$, $b$ and $c$ denote values obtained in this work, in Ref.\ \citenum{tio2ref3prb} and in Ref.\  \citenum{tio2ref3exp}, respectively.}
    \label{tab:bulk}
\end{table}
The calculated surface energies are listed in Table \ref{tab:anasurf}.  The current parameterization slightly overestimates the surface energies for anatase (001) and rutile (110) compared to PBE and PBEsol results. The anatase (101) and the rutile (001) and (100) surfaces are underestimated compared to PBEsol results and overestimated compared to PBE. The energetic ordering of the surfaces with respect to each other is reproduced nicely within the \textit{tio2nano} set. Overall the current set shows a good agreement to ab initio methods.
\begin{table}[h]
    \centering
    \begin{tabular}{c c c c c c}
        \hline
         & DFTB$^a$ & PBE$^a$ & PBEsol$^a$ & PBE0$^b$ \\
        \hline
        anatase & & & &\\
        (001)  & 1.36 & 0.95 & 1.23 &  1.25* \\
        (101)  & 0.49 & 0.45 & 0.80 &  0.58\\
        & & & & \\
        rutile & & &  \\
        (100) & 1.07 & 0.62 & 1.11 & 0.83 \\
        (110) & 0.97 & 0.37 & 0.88 & 0.46 \\
        (001) & 1.47 & 1.33 & 1.79 & 1.59\\
        \hline
    \end{tabular}
    \caption{Calculated anatase and rutile surface energies in J/m$^2$ for selected surfaces. The index $a$ denotes  values obtained in this work, while values marked with $b$ were taken from Ref. \citenum{tio2ref1}, except the one indicated by *, which was taken from Ref.~\citenum{tio2ref2}.}
    \label{tab:anasurf}
\end{table}

\begin{figure}[htbp]
    \centering
    \includegraphics[width=0.45\textwidth]{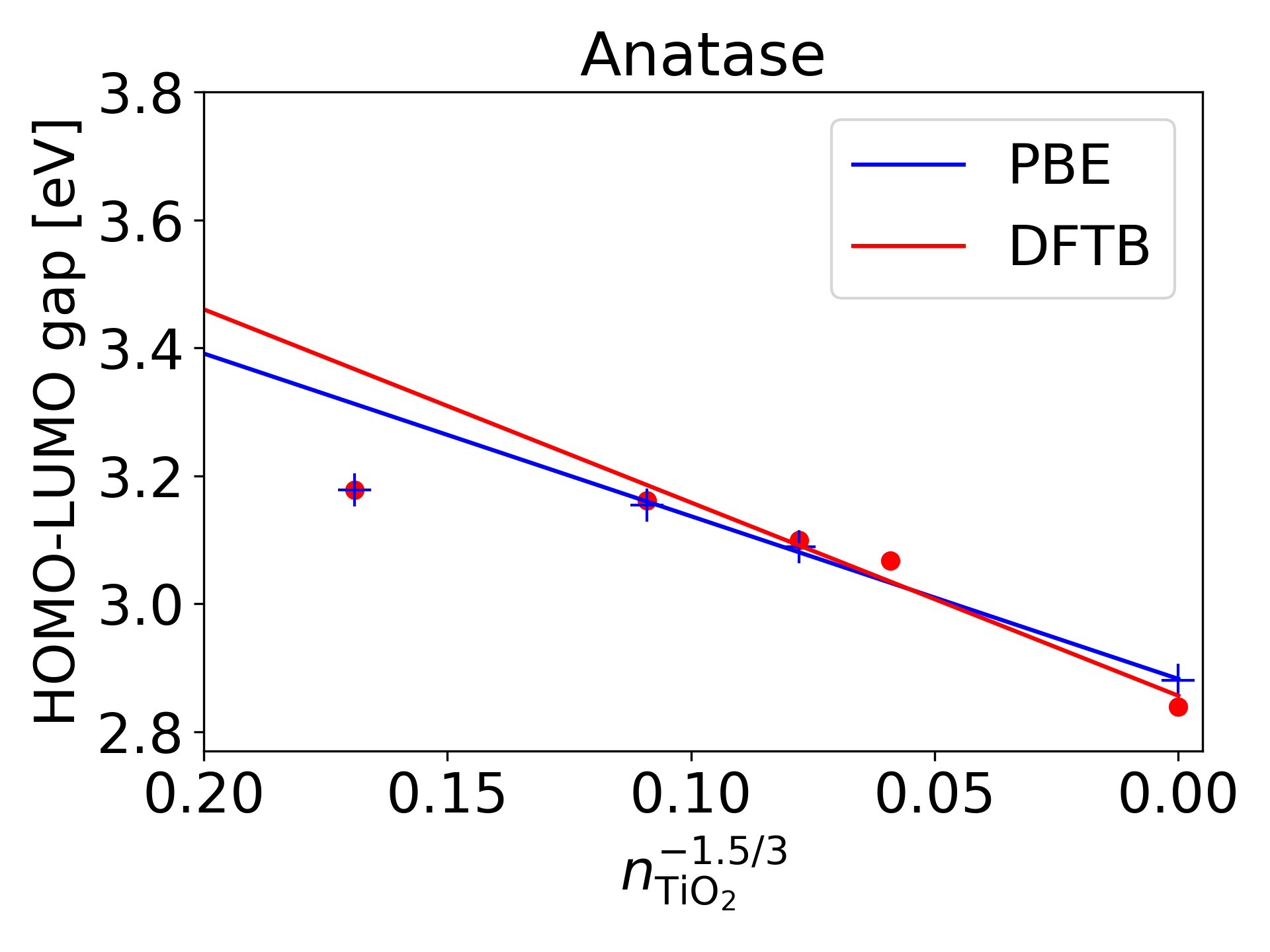}
    \includegraphics[width=0.45\textwidth]{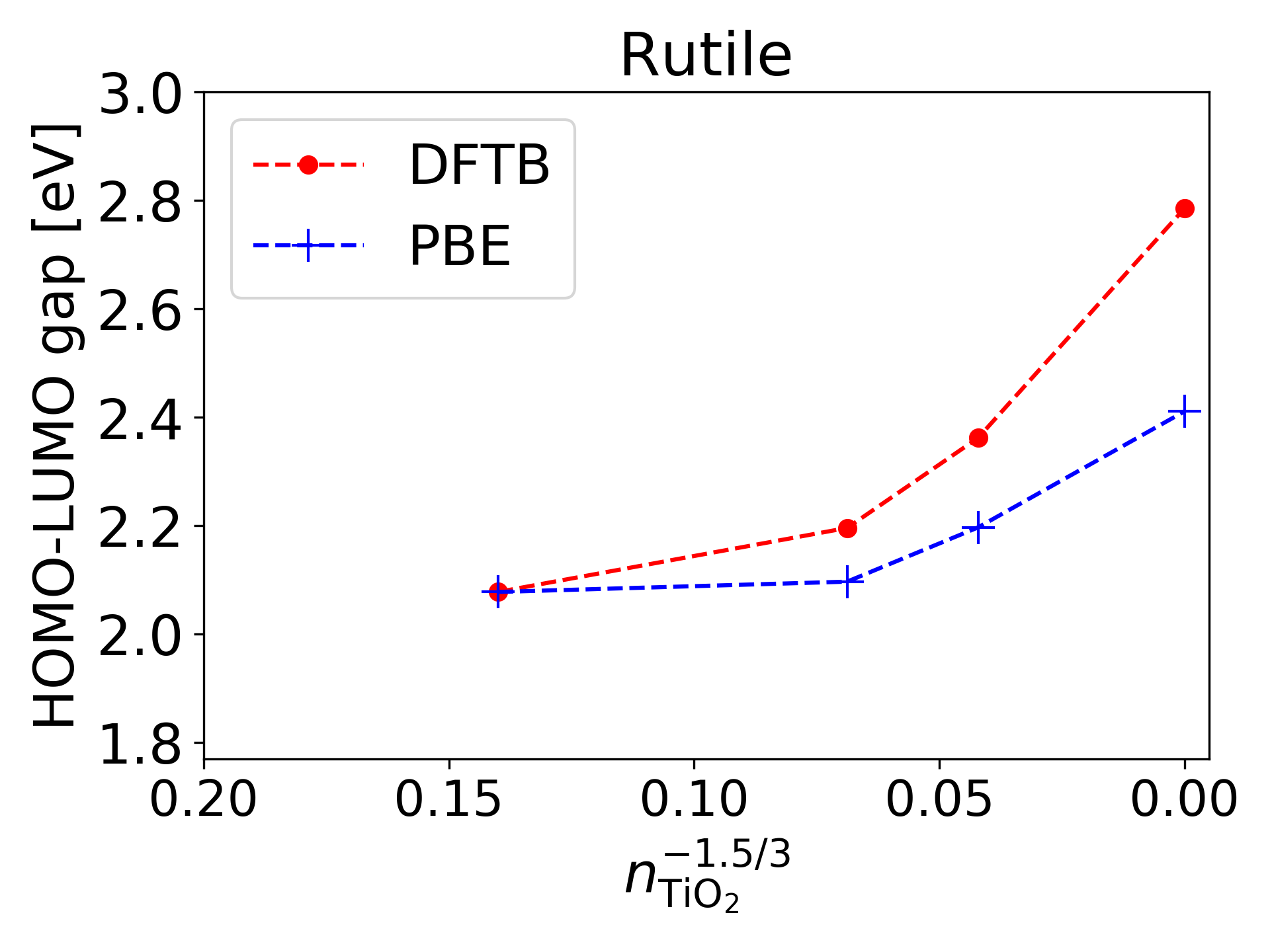}
    \caption{Gap energy behavior over particle size for anatase (left) and rutile (right). The DFTB values were obtained with the new parameter set and are displayed in addition to the PBE results. The PBE results were aligned with the DFTB ones to match the value for the smallest nanocrystal structure. Note that for anatase the smallest crystal size was not included in the linear fit as that structure did not follow the quantum confinement behaviour for the gap states. The straight solid lines represent a linear fit, while the dashed lines serve only as guides to the eyes.}
    \label{fig:ana-rut-gap-full}
\end{figure}

\begin{figure}[htbp]
    \centering
    \includegraphics[width=0.8\textwidth]{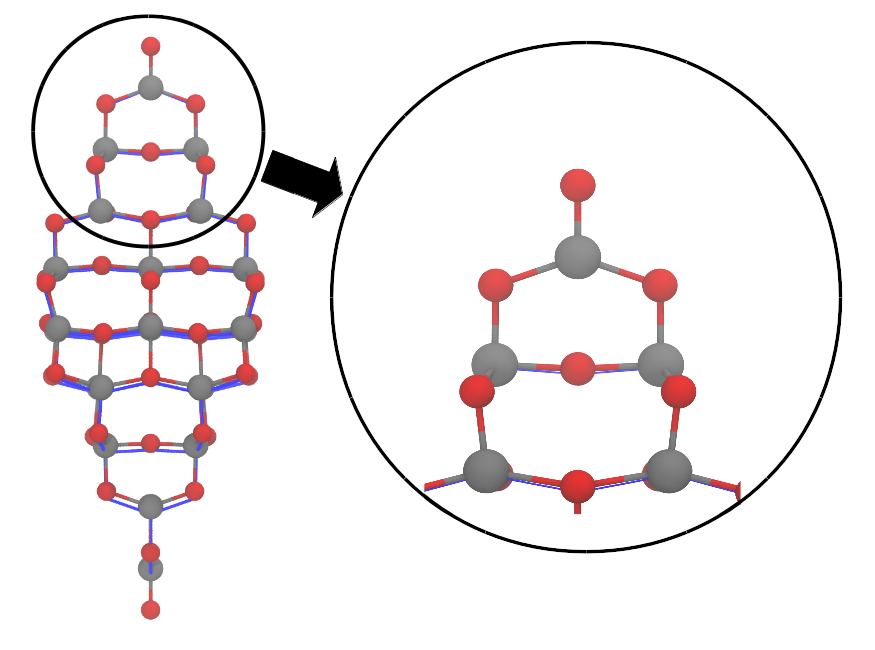}
    \caption{PBE Ti$_{35}$O$_{70}$ relaxed nanocrystal displayed by blue lines aligned at the top oxygen atom to the DFTB relaxation with the new parameter set (shown by sticks and balls). Relaxation at the tip is so close to the PBE counter part that the blue lines are almost not visible in the representation.}
    \label{fig:rightrelax}
\end{figure}

Figure \ref{fig:ana-rut-gap-full} shows the PBE gap values of anatase and rutile in comparison to the ones obtained with the newly developed parameter set. The behavior of both sets is now comparable to Ref.~\citenum{nanomain}. The smallest anatase nanocrystal considered shows with the ChIMES based repulsive function a relaxation which is very close to the one obtained with the PBE calculation (see Fig.~\ref{fig:rightrelax}). Overall, the new parameters reasonably describe the various forms and phases of titania. The band alignments between the phases are described in the following section.

\subsection{Band alignment}

The primary aim of the current paper is to establish a computationally feasible approach that allows us to study band alignment across a TiO$_2$ anatase-rutile nano-particle interface and to determine the role of the interface in this alignment. The approach is based on the customized DFTB parametrization described above. The quality of the new method is assessed by comparison to calculations at the DFT-PBE level of theory.

First we present the bulk band alignment between anatase and rutile using the branching point energy approach as it was done for TiO$_2$ in Ref.~\citenum{PeterBA} previously and compare the PBE and the DFTB values. Afterwards, following Ref.~\citenum{nanomain}, the level alignment deduced from non-bonded nano-particle structures will be discussed. We compare previous results to our DFT-PBE calculations as well as to results obtained with DFTB. Finally, we investigate the influence of the interface on the band alignments by DFTB-calculations on bonded nano-particles.

\subsubsection{Bulk band alignment}
\label{SEC:BulkBA}

\begin{figure}[htbp]
    \centering
    \includegraphics[width=0.6\textwidth]{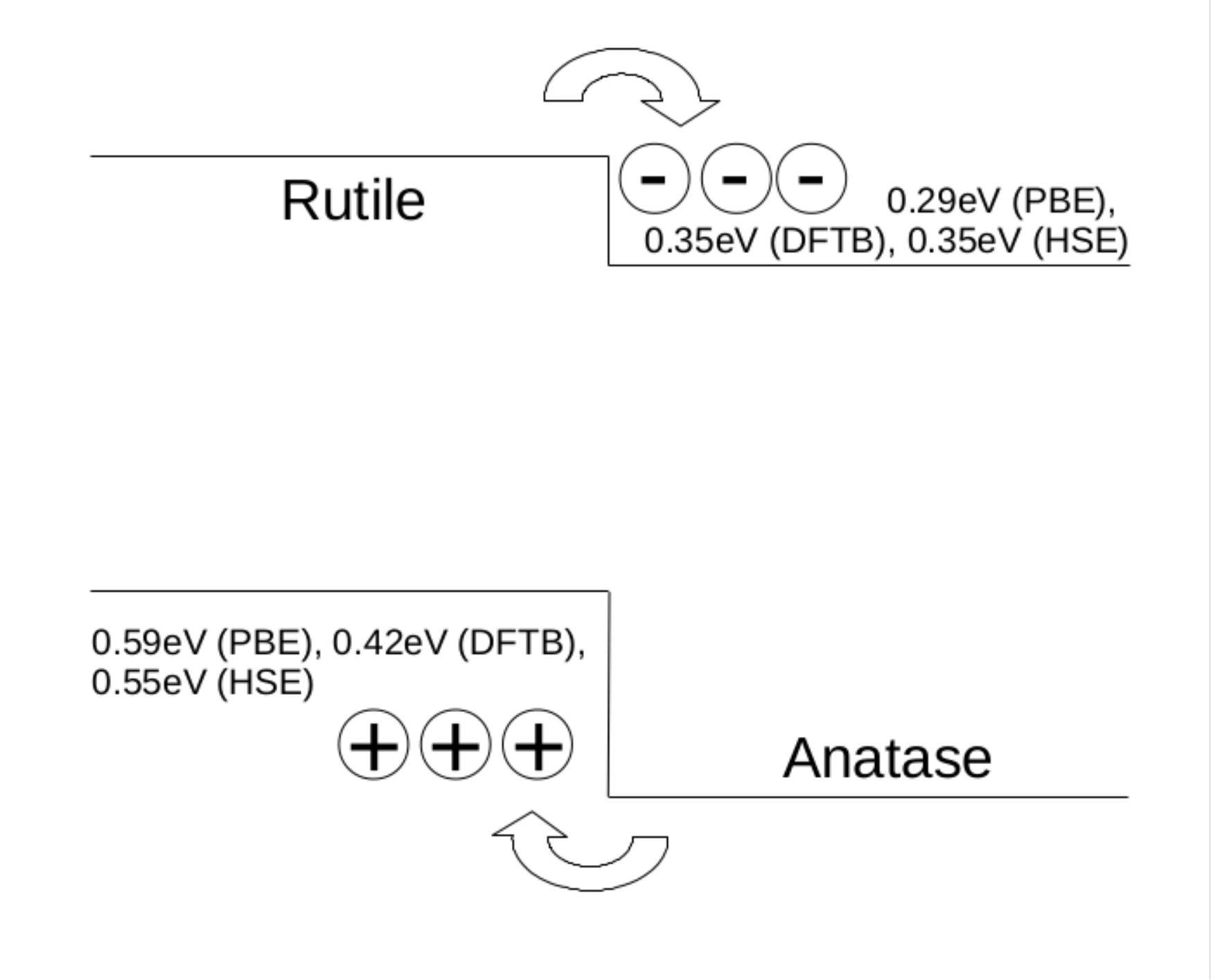}
    \caption{Anatase and rutile band alignments computed by the branching point technique with DFTB (\textit{tio2nano}), PBE and HSE06. The HSE06 values had been taken from Ref.~\citenum{PeterBA}.}
    \label{fig:align-bulk}
\end{figure}

First we consider the band alignment between the bulk phases using the branching point technique. The branching point energy $E_\mathrm{BP}$, or charge neutrality level, can be computed from the average of the mid-level states using the relation
\begin{align}
    E_\mathrm{BP} = \frac{1}{2N_\mathbf{k}} \sum_\mathbf{k} \left[ \frac{1}{N_\mathrm{CB}} \sum_i^{N_\mathrm{CB}} \epsilon_\mathrm{CB}^{\mathbf{k},i} + \frac{1}{N_\mathrm{VB}} \sum_i^{N_\mathrm{VB}} \epsilon_\mathrm{VB}^{\mathbf{k},i} \right]
    \text.
\end{align}
The index $\mathbf{k}$ runs over the $N_\mathbf{k}$ k-points being considered. In every k-point $N_\mathrm{VB}$ valence band and $N_\mathrm{CB}$ conduction band states with respective eigenvalues $\epsilon_\mathrm{CB}^{\mathbf{k},i}$ and $\epsilon_\mathrm{VB}^{\mathbf{k},i}$ are averaged. We chose $N_\mathrm{VB}=1$ and $N_\mathrm{CB}=1$ independently of possible degeneracies of the bands, as the alignment is known to depend only weakly on the degeneracy and on the number of included bands \cite{PeterBA}. Figure \ref{fig:align-bulk} compares our results for E$_\mathrm{BP}$ to those in Ref.~\citenum{PeterBA}. The results obtained with the \textit{tio2nano} set compare well to the calculated PBE values as well as to the HSE06 values in Ref.~\citenum{PeterBA}. All levels of theory predict that rutile and anatase act as a hole and as an electron trap, respectively.

\subsubsection{Band alignment in non-bonded nanocrystal systems }\label{sec:band-nano}

Following Ref.~\citenum{nanomain} we performed DFT-PBE and DFTB calculations for anatase and rutile nanocrystals of different sizes. We have plotted the HOMO and LUMO levels as well as their gap values as function of the number of TiO$_2$ formula units in the structure raised to the factor ${-\alpha/3}$ . The parameter $\alpha$ had been set to 1.5 as this yielded a better fit to our data than the value of 1.35 used in Ref.~\citenum{nanomain}. Note, that we did not include the smallest anatase crystal in the linear fit, as we found it to not follow the particle in the box model (neither for PBE nor for DFTB). The PBE and DFTB results for anatase are in good agreement with the PBEx data of Ref.~\citenum{nanomain}. By extrapolating the DFTB nanocrystal results to infinite particle sizes using a linear fit with respect to $n^{-1.5/3}$, we obtained a band gap of $3.02$~eV as compared to the bulk value of $2.84$~eV. A similar overestimation can be seen in the extrapolation of the PBE data, which yields a gap of $2.25$~eV as compared to the PBE bulk value of $2.09$~eV. 

\begin{figure}[htbp]
    \centering
    \includegraphics[width=0.45\textwidth]{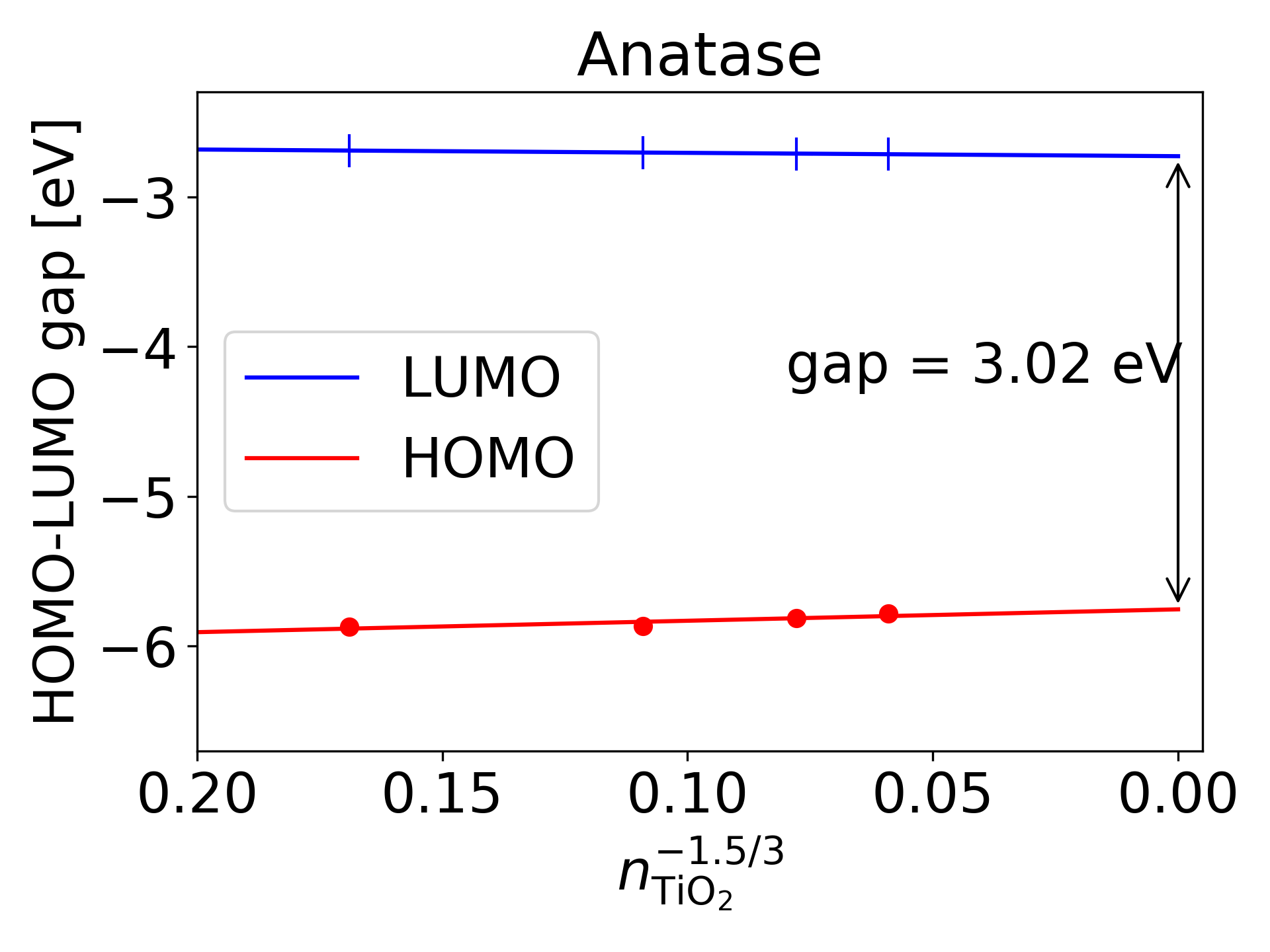}
    \includegraphics[width=0.45\textwidth]{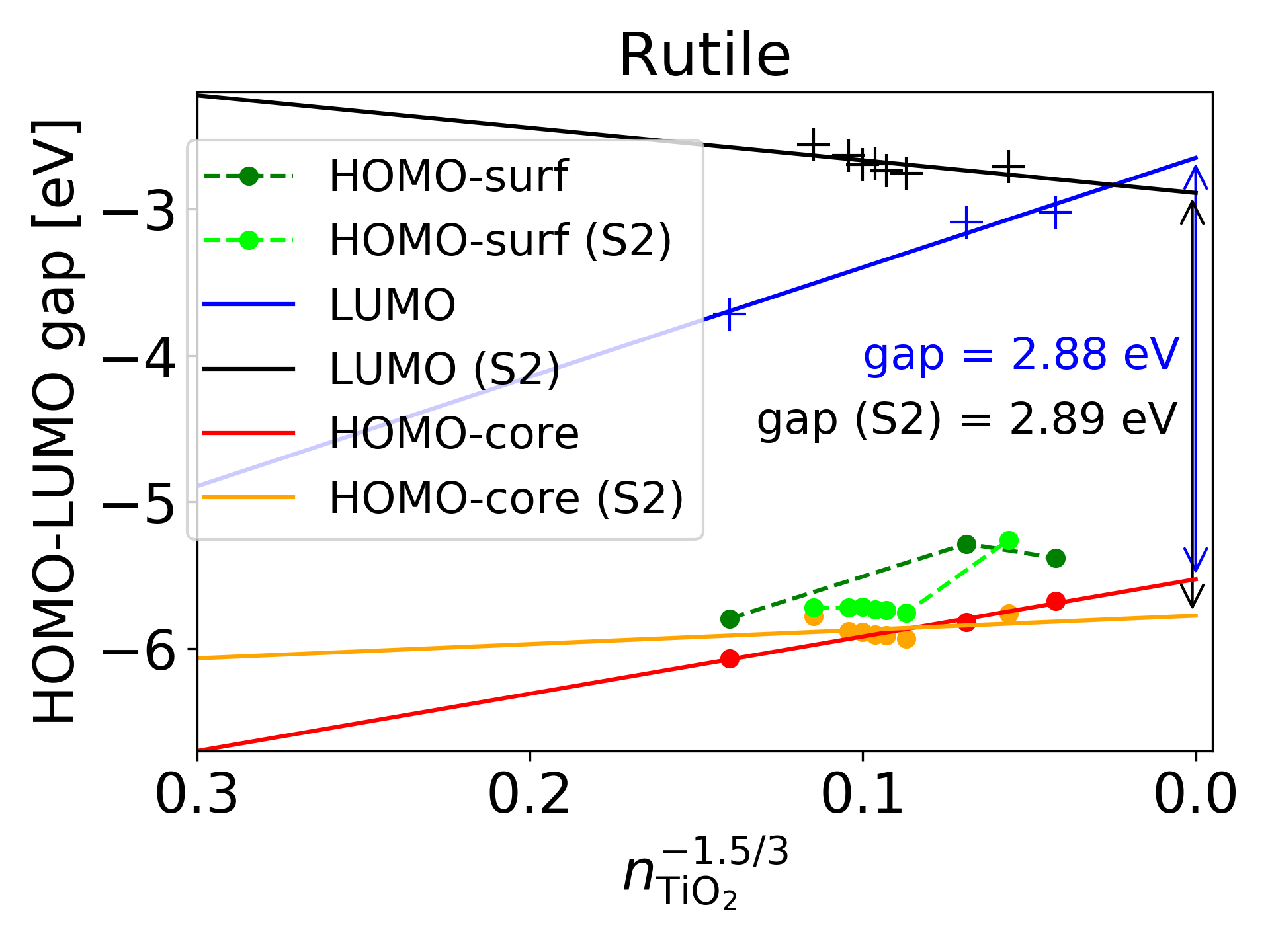}
    \caption{HOMO and LUMO levels for anatase (left) and rutile (right) calculated with DFTB. For rutile both, the highest occupied core states (HOMO-core) and the occupied surfaces states (HOMO-surf) are shown. The corresponding levels for the second rutile crystal set are indicated with (S2). The dashed connection lines serve only as guides to the eye. The indicated gap sizes of the bulk phases are obtained from the linear fits of the respective HOMO and LUMO states.}
    \label{fig:ana-rut-gap}
\end{figure}
\begin{figure}[htbp]
    \centering
    \includegraphics[width=0.45\textwidth]{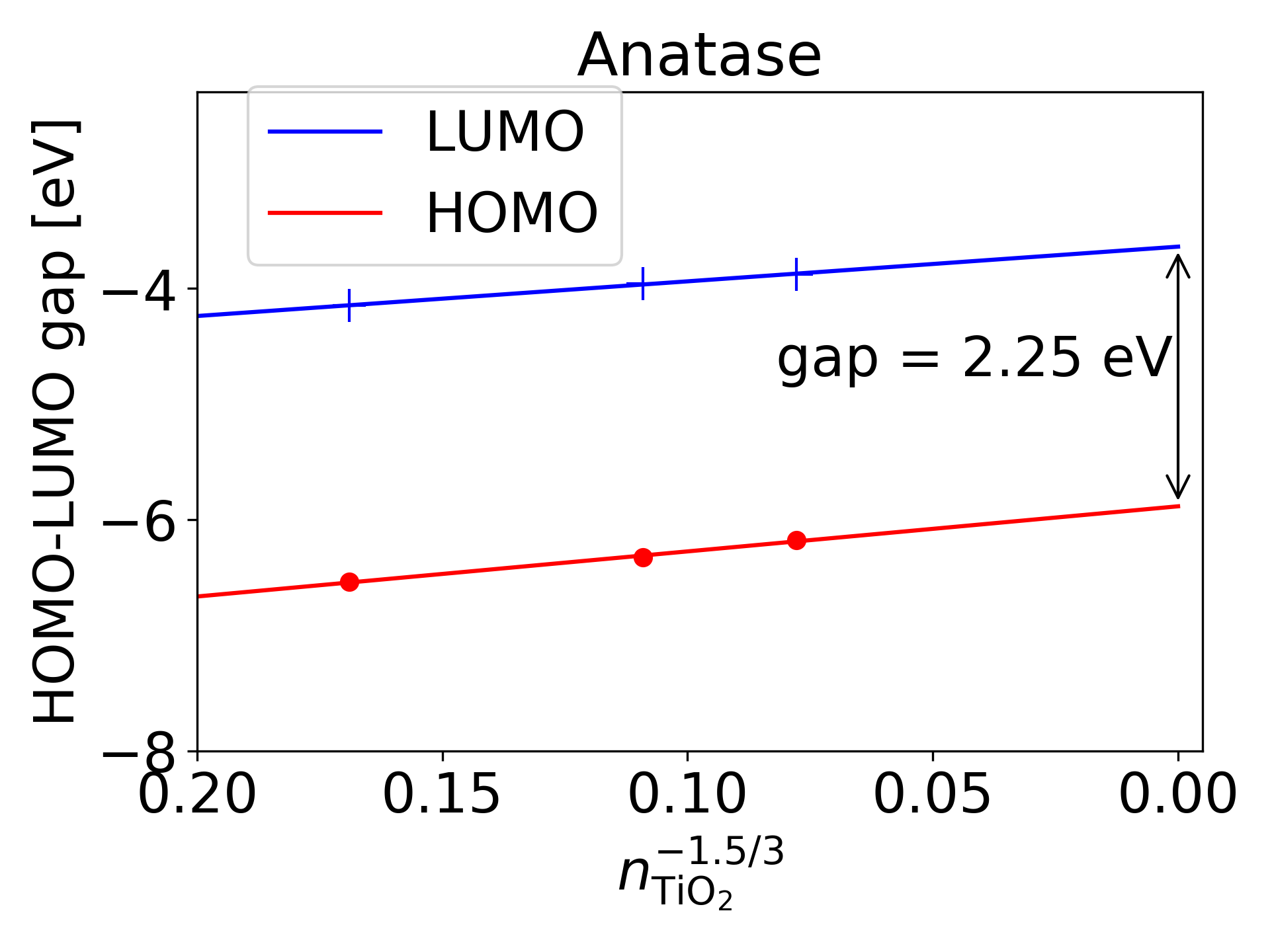}
    \includegraphics[width=0.45\textwidth]{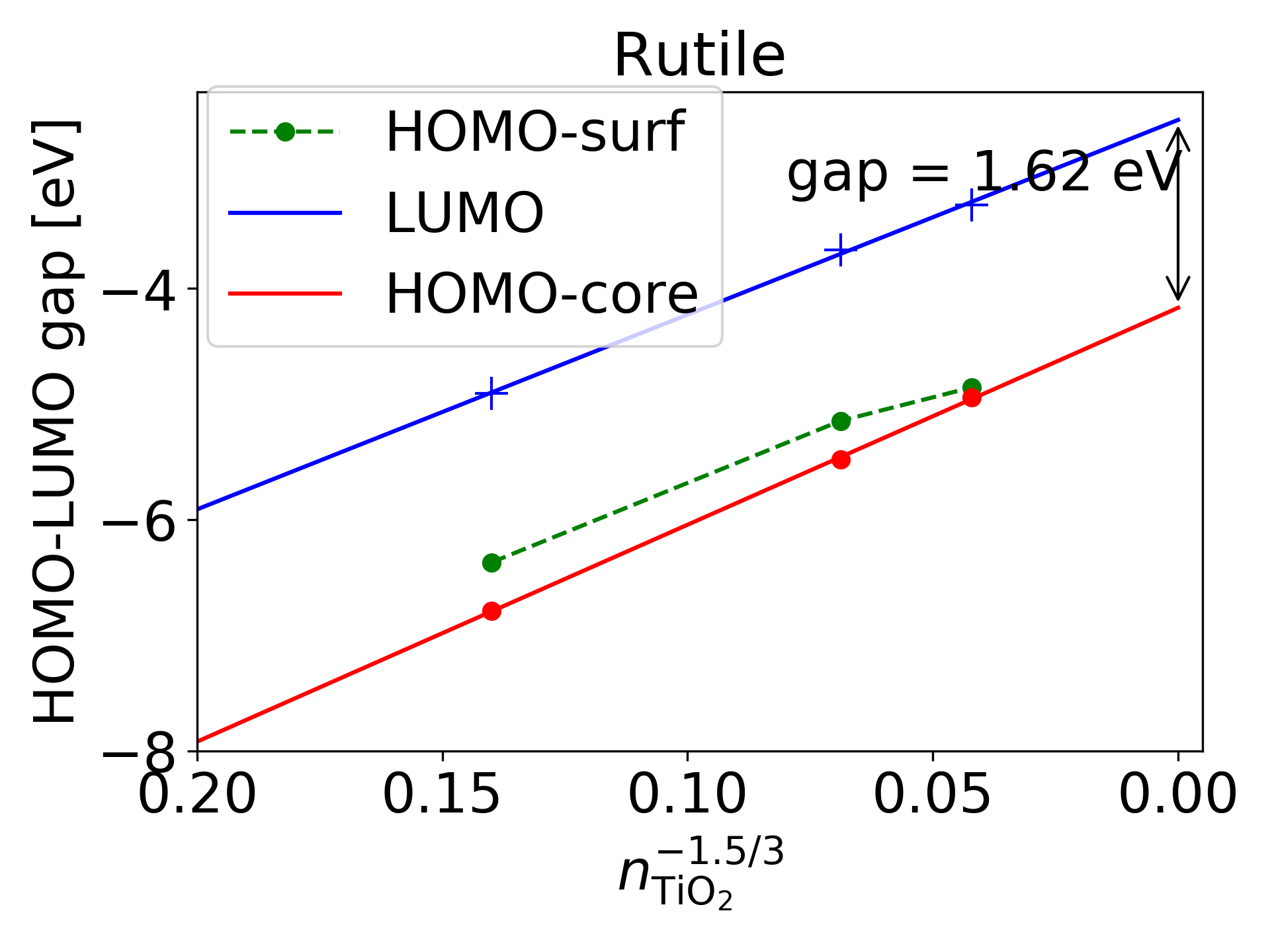}
    \caption{HOMO and LUMO levels for anatase (left) and rutile (right) calculated with PBE. For rutile both, the highest occupied core states (HOMO-core) and the occupied surfaces states (HOMO-surf) are shown. The dashed connection lines serve only as guides to the eye. The indicated gap sizes of the bulk phases are obtained from the linear fits of the respective HOMO and LUMO states.}
    \label{fig:ana-rut-gap-PBE}
\end{figure}

In the case of rutile nanocrystals, extrapolation of the band gap is complicated by the presence of surface states. This leads to a non-linearity of the band gap caused by a non-linear behavior of the VB-edge. As shown in Fig.~\ref{fig:ana-rut-gap}, although the highest lying occupied core-states clearly follow the confinement linearly, an occupied surface state with non-linear behavior appears in the gap. This leads to an initial narrowing of the gap with the particle size before reaching the point where the surface and core states cross and after which the gap follows a linear behaviour again. A similar behavior can be found in the PBE-results as shown in Fig.~\ref{fig:ana-rut-gap-PBE}. The observed behavior of the surface states indicates that the simple text-book particle in a box model can be applied only to core-states but not to surface states. 

Comparing the extrapolated value of the band gap for rutile, excluding the surface states, from the nanocrystals with the calculated band gap of the bulk, we find that the extrapolated value slightly overestimates the bulk one with DFTB (2.88~eV vs.\ 2.76~eV) and underestimates it with PBE (1.62~eV vs 1.79~eV). We also note that the rutile data indicate an inverse quantum confinement trend even when excluding the surface states, in disagreement with the PBEx data presented in Ref.~\citenum{nanomain}. We are not entirely sure what the origin of the discrepancy is. We note however, that both rutile slabs \citenum{Bredow} as well as nanowires \citenum{Hmiel} exclusively exposing (110) surfaces show an even-odd behaviour in the valence and conduction band edges with number of layers. For odd numbers, both display an inverse quantum confinement behaviour, while for even numbers they display regular quantum confinement behaviour. The effect was found to be more pronounced for the conduction band states.

\begin{figure}[htbp]
    \centering
    \includegraphics[width=0.45\textwidth]{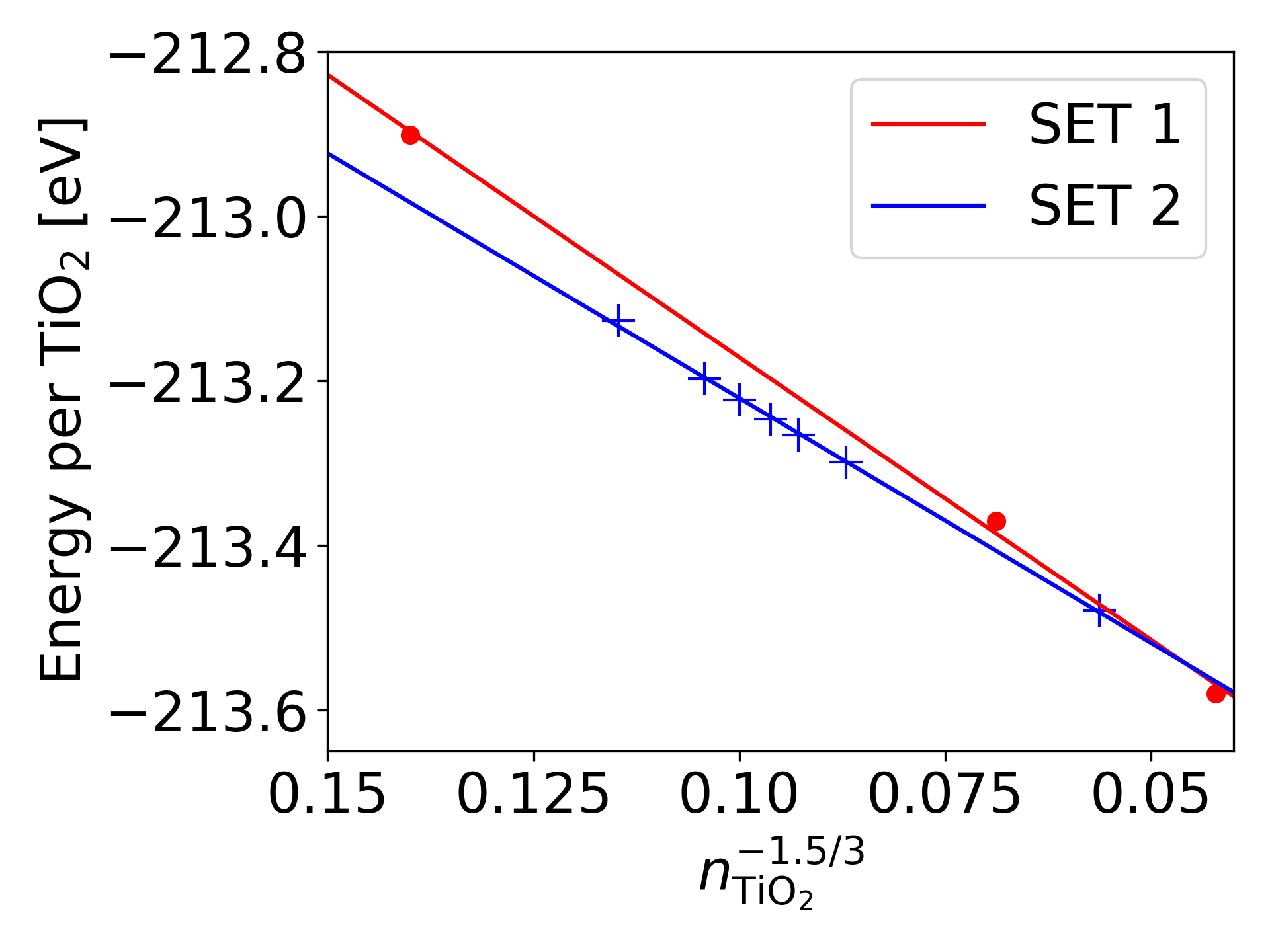}
    \caption{Comparison of stability of rutile nanocrystal in the orginal set from Ref.~\citenum{nanomain} (SET 1) and the new set (SET 2) as a function of size. Energies are given per TiO$_2$ formula units.}
    \label{fig:new-rutile-stab}
\end{figure}

Therefore, in addition, we also considered another set of quasi Wulff-type rutile nanocrystals by cutting particles along the (110) and (101) planes with an even number of (110) layers across the waist of the particle. In the following we will refer to this new set as SET 2. These nanocrystals contain 76, 92, 100, 108, 116, 132 and 316 TiO$_2$ formula units. In order to make the crystals stoichiometric, we removed a number of Ti atoms at the (101) facets as opposed to adding dangling oxygen ions as was done in Ref.~\citenum{nanomain}. The particles of this set are found to be more stable compared to those presented in Ref.~\citenum{nanomain}, see Fig. \ref{fig:new-rutile-stab}. We also tried building larger particles with even numbers of (110) layers, but these particles became metallic and also became significantly less stable. Using the stable and non-metallic particles of the new set we obtain an extrapolated value (excluding surface states) for the band gap of 2.89 eV, see Fig.~\ref{fig:ana-rut-gap}. We note that the new set displays a regular quantum confinement trend with a decreasing HOMO-core to LUMO gap with increasing crystal size.  While the HOMO-core states of the rutile nanocrystals of the new set are located in the same energy region as those of the original set, the LUMO states are located at higher energies. This behaviour is therefore consistent with the even-odd behaviour with respect to the number of (110) layers for the VB and CB positions for slabs and nanowires in Refs. \citenum{Bredow} and \citenum{Hmiel}, respectively.

Given that particles of comparable stability may show contrasting behaviour in terms of their HOMO and LUMO positions suggests that we need to be careful in establishing scaling relations -- explicit simulations are clearly required for small rutile nanocrystals. 

Using the HOMO and LUMO level positions obtained above, band level alignments in combined nanocrystals, consisting of a rutile and an anatase nanocrystal, can be predicted. Note, however, that such predictions would not take the effect of the actual anatase-rutile interfaces into account which can have a significant impact as will become clear from the presentation in the section on \textit{Band alignment in bonded nanocrystals}. We also note that much of such size dependence in the alignment will be driven by the large variation in the HOMO and LUMO position in the rutile nanocrystals. The results will also depend on the type of rutile nanocrystals we use.  Nevertheless, combining an anatase nanocrystal with rutile nanocrystals of type presented in Ref.~\citenum{nanomain}, one obtains a so-called type II rutile alignment (using the nomenclature of Ref.~\citenum{nanomain}) for small rutile nanocrystals, where rutile acts as an electron trap and anatase as a hole trap. Increasing the size of the rutile nanocrystal results in a transition to a so-called type I alignment, where both the HOMO and the LUMO states are located within the rutile nanocrystal. Finally, for even larger rutile crystal sizes one obtains a so-called type II anatase alignment (as also predicted by the branching point calculation with the bulk structures), where rutile acts as a hole and anatase as an electron trap. The same quantitative behaviour was derived for PBEx in Ref.~\citenum{nanomain}.

On the other hand, combining an anatase nanocrystal with a rutile nanocrystal from the new set (S2) one can obtain a number of different alignments. Most notably, using the smallest rutile nanocrystal of the new set leads to an alignment where both the HOMO and the LUMO states are located within the anatase nanocrystal. 

With the aforementioned difficulties in establishing robust trends in the HOMO and LUMO levels in rutile nanocrystals, we conclude that we are unable to make robust prediction in HOMO-LUMO level alignment between anatase and rutile nanocrystals, at least for size range covered by our simulations, $\sim$ 1-4.5 nm, without resorting to explicit simulations.

\subsubsection{Band alignment in bonded nanocrystals}\label{sec:InterBA}

\begin{figure}[htbp]
    \includegraphics[width=0.4\textwidth]{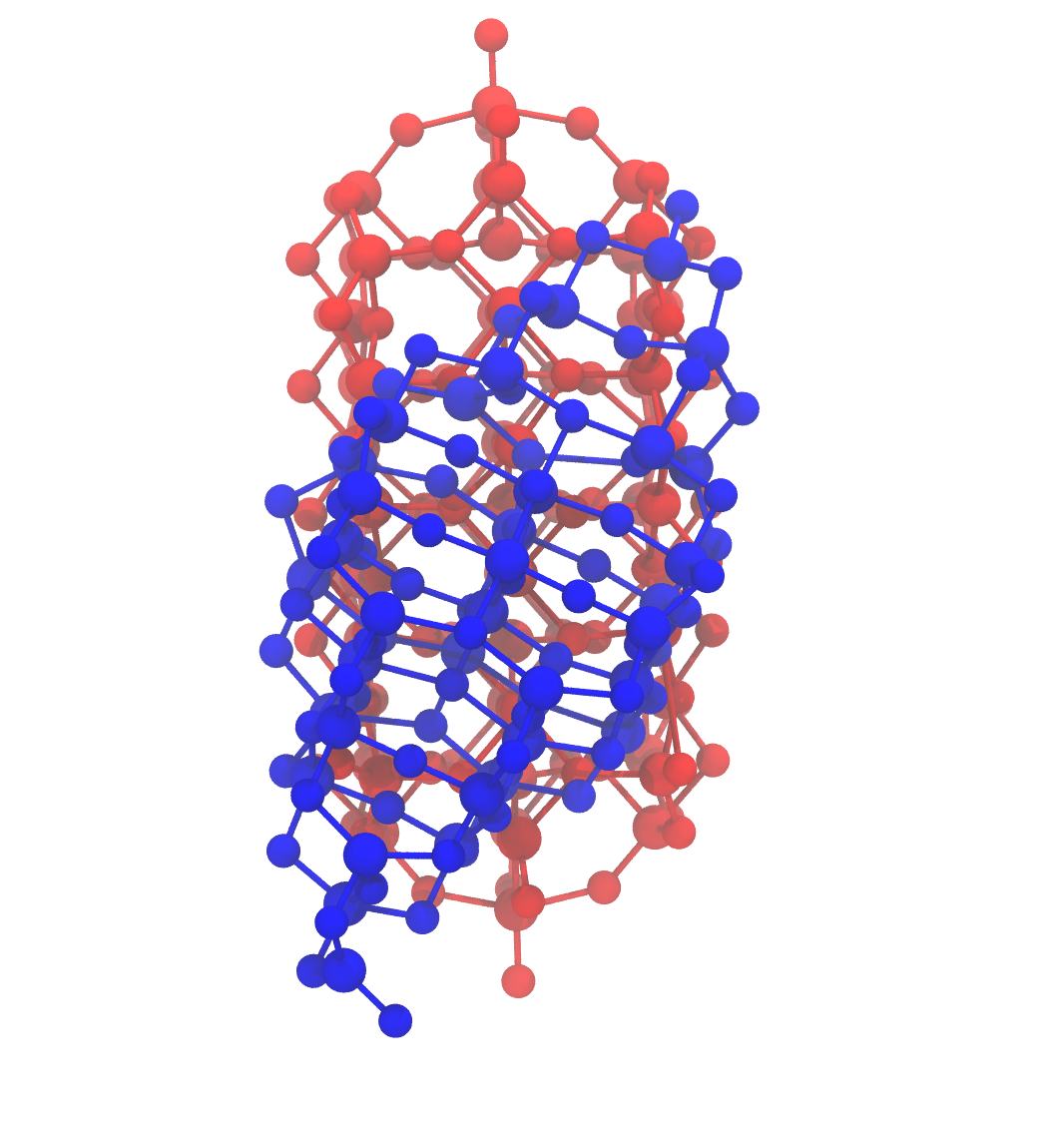}
    \includegraphics[width=0.4\textwidth]{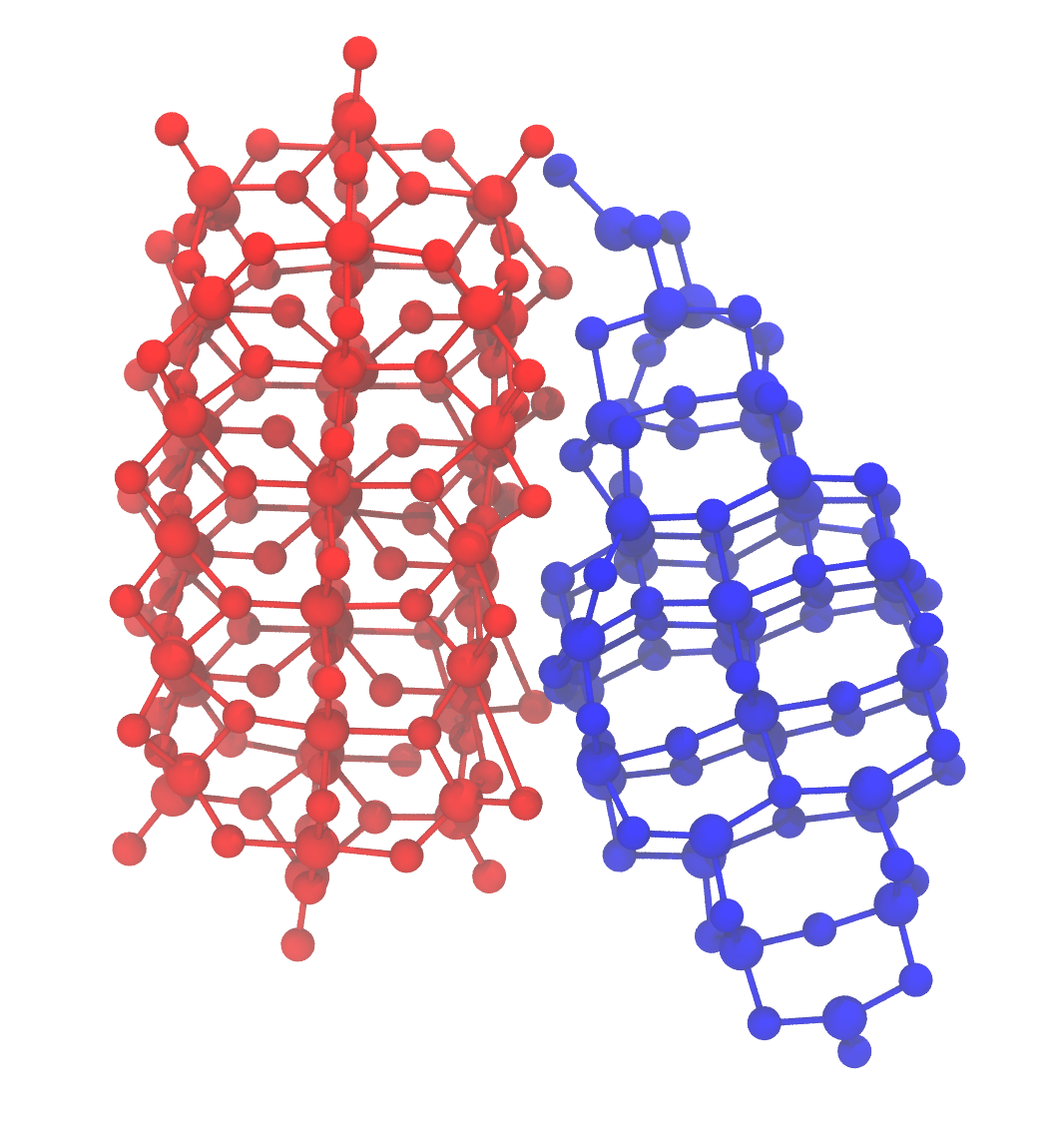}
    \caption{Typical alignment of the most stable rutile-anatase nanocrystal interface structures. The red and blue structures represent the rutile and anatase nanocrystals, respectively.}
    \label{fig:ana-rut-stable}
\end{figure}

Exploiting the efficiency of the DFTB method, we have investigated the influence of the interface on the gap alignment by simulating explicit interface structures between different orientations of the smallest anatase and rutile nanocrystals. In the most stable structures we have obtained for nanocrystals with a rutile (110) and anatase (101) interface,  the nanocrystals tend to align with an angle of around 30 degrees with respect to each other along the rutile [001] and anatase [010] directions. This alignment, as shown in Figure~\ref{fig:ana-rut-stable}, maximizes the interaction area and the bond formation between the two structures, minimizing also the interface energy which was computed according to Equation~\ref{eq:einter}.

\begin{figure}[htbp]
    \centering
    \includegraphics[width=0.7\textwidth]{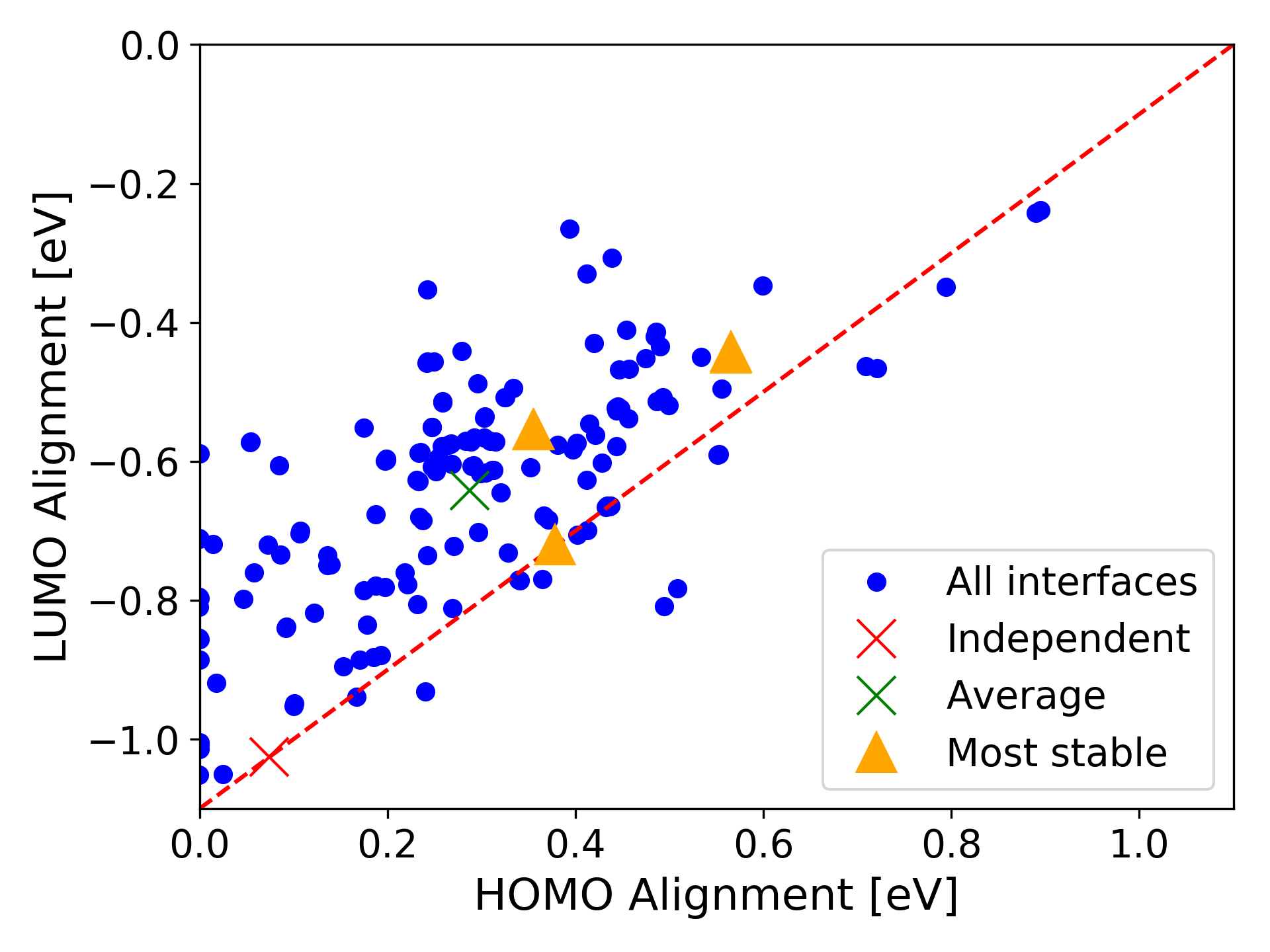}
    \caption{HOMO and LUMO alignment of TiO$_2$ nanoparticles composed of rutile and anatase nanocrystals. The alignment is positive if the respective edge state in rutile is higher as in anatase. Blue dots indicate values obtained from bonded nanocrystals, green crosses their average, and the red cross the value obtained from the non-interacting nanocrystals. The five most stable structures are marked with yellow triangles, but since two pairs have very similar alignment values, only three triangles can be seen. The dashed diagonal line serves a guide for the eye and moving parallel to this line correspond to alignments with constant gap sizes for both nanocrystals. In the line shown, the values for these gaps are  2.08~eV for rutile and 3.18~eV for anatase and corespond to the HOMO-LUMO gaps of the isolated particles.}
    \label{fig:aligncbvb}
\end{figure}

The prediction using independent nanocrystals (neglecting the interface effects) suggests a type I alignment with a mismatch of 0.07~eV for the HOMO and 1.03~eV for the LUMO. This suggests a complete domination of the gap by rutile states, due to the narrowing of its gap by surface states. The gap alignments for all 156 investigated bonded arrangements, their average and the value obtained from the non-interacting nanocrystals are displayed in Figure~\ref{fig:aligncbvb}. The 5 configurations with the lowest energy has been also marked in the figure. It is obvious that although the prediction using non-interacting nanocrystals delivers the correct alignment type, the magnitude in the offset are far off from both the average interface offset and the one obtained from the most stable interface structures. This suggests, that interface effects have a significant influence on the level alignment and must be accounted for explicitly. One observes a slight correlation of the LUMO/HOMO alignments along with the indicated diagonal line which would correspond to rigid shifts of the bands edges without any change in band gap. This therefore represents situations, where bands in both phases are shifted due to an interface dipole that varies in magnitude depending on the relative orientation of the nanocrystals, but where no new interface states are being created. 

\begin{figure}[htbp]
    \centering
    \includegraphics[width=0.3\textwidth]{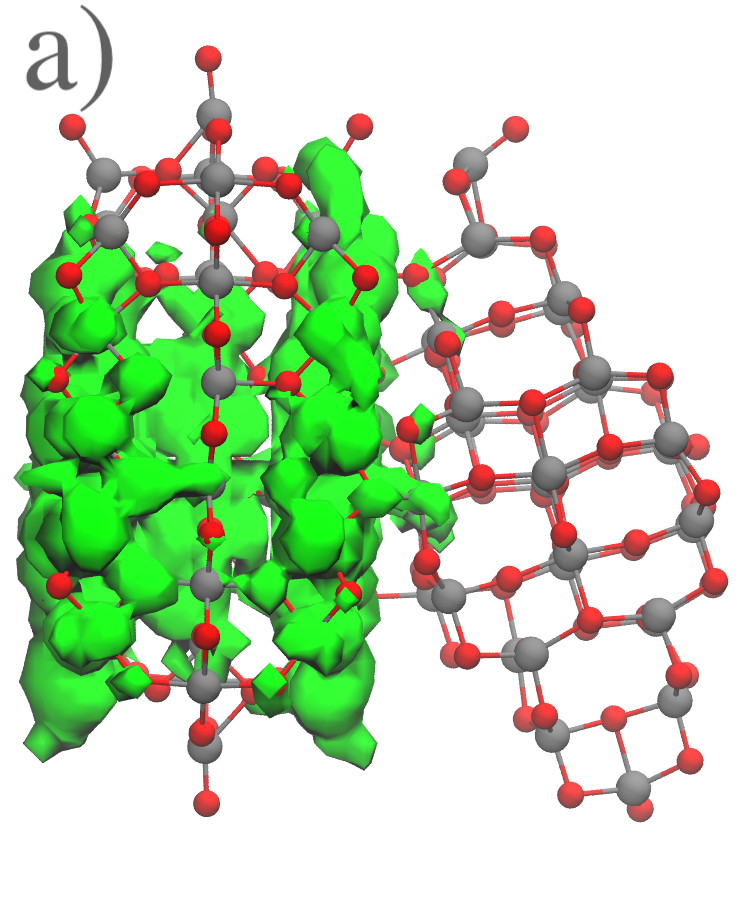}
    \includegraphics[width=0.3\textwidth]{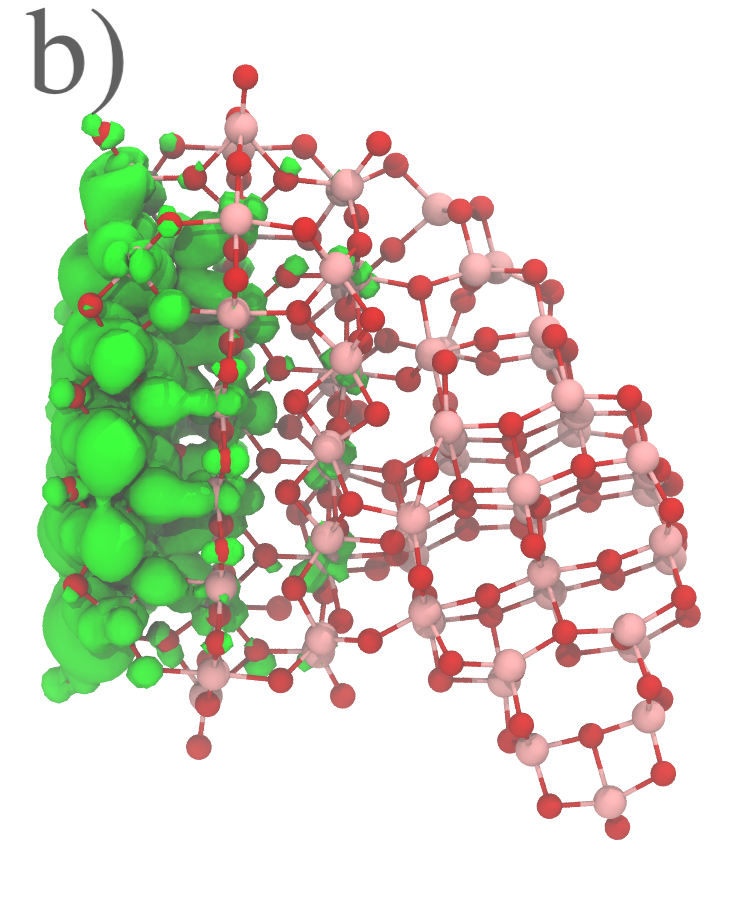}\\
    \includegraphics[width=0.3\textwidth]{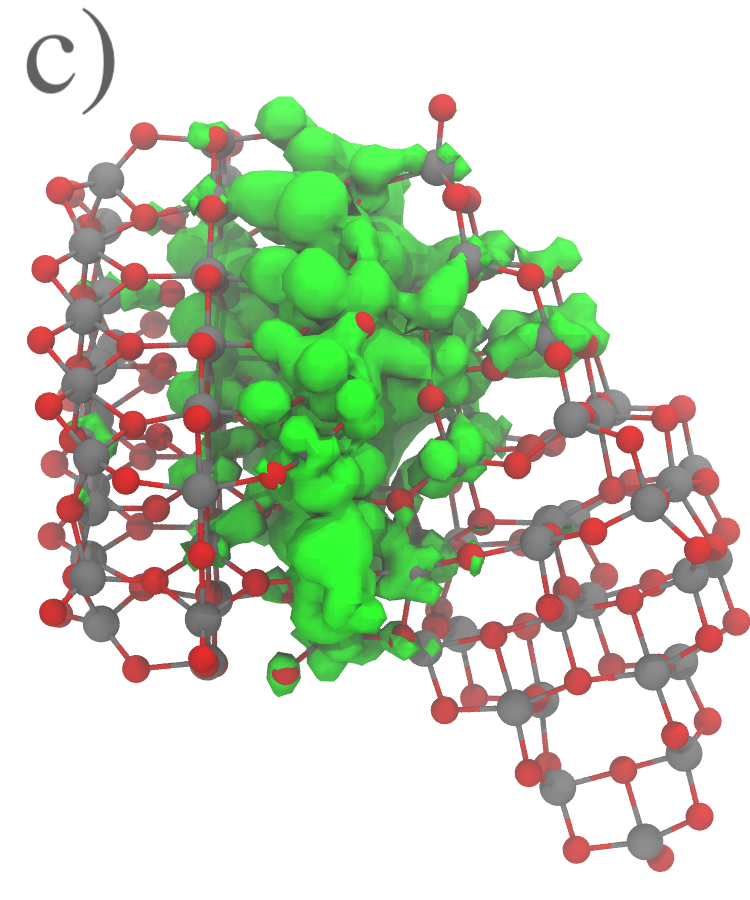}
    \includegraphics[width=0.3\textwidth]{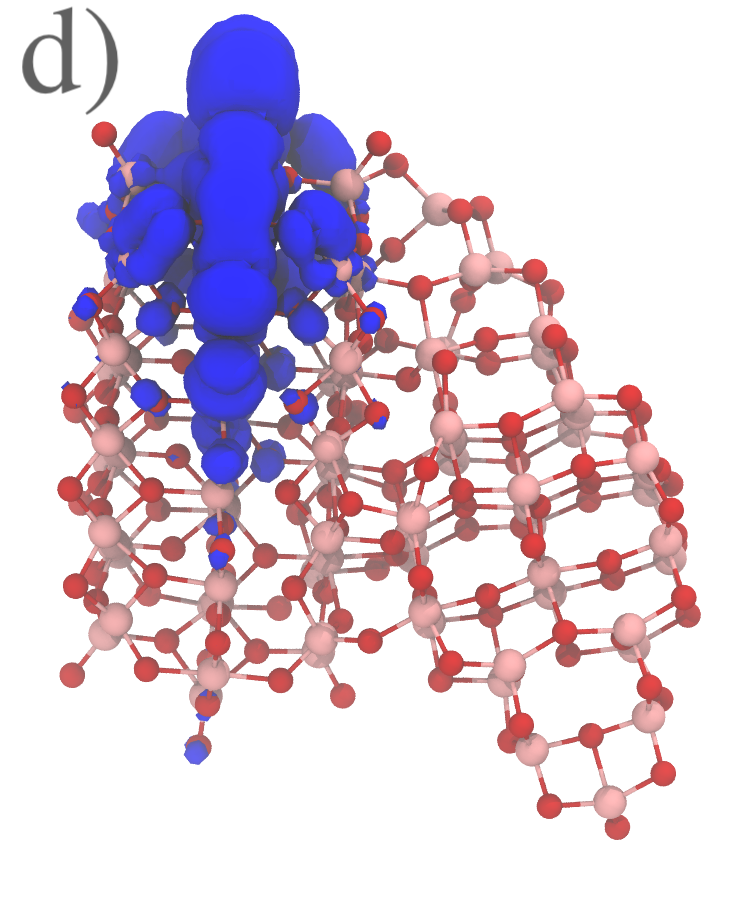}
    \caption{Typical LUMO and HOMO wavefunction in the anatase-rutile nanocrystals.
    Subfigure a) shows the least common LUMO wave function type which is similar to the LUMO of an isolated rutile nanocrystal, b) shows the most common interface wave function, where wave function is not present in the interface region, c) shows the second most common interface type, where the LUMO wave function is located at the interface. Subfigure d) displays the HOMO wave function, which is similar in all interfaces and corresponds to the HOMO-surface state of the isolated rutile nanoparticles.}
    \label{fig:wave-functions}
\end{figure}

Some typical LUMO and HOMO wave functions of the interfaces are displayed in Fig.~\ref{fig:wave-functions}.  We found three different types of LUMO wave functions and one HOMO wave function. The HOMO wave function is identical to the rutile HOMO-surface state. The most common LUMO wave function is located exclusively in the rutile nanoparticle, but differs considerably from the LUMO wave function of the isolated rutile nanocrystal. The second most common is located directly at the interface, while the least common one, which only occurs in four of the investigated cases, resembles the original rutile LUMO wave function. The interface in each of the five most stable structures is of the first type, which is also the most common among all investigated nanoparticles. This underlines the necessity to model nanocrystal interfaces explicitely, as the LUMO edge states might differ considerably from the those obtained from the independent particle model. This aspect would be important to consider not least, when modeling catalytic reactions at the surface of nanocrystals.

\section{Conclusions}

We have developed a DFTB parameter set using the ChIMES force field with three-body terms as repulsive potentials especially designed for rutile and anatase nanocrystals. We found that the three-body term was crucial in order to predict the correct relaxation around the low coordinated atoms at the apex of anatase nanocrystals. 

We used the new DFTB parametrization, the \textit{tio2nano} set, to investigate different anatase-rutile band alignment models in order to predict effective charge carrier separation in mixed anatase-rutile systems. While the bulk band alignment model predicts a type II anatase alignment where anatase acts as an electron and rutile as a hole trap in accordance with the literature \cite{PeterBA,slabBA,baexp}, the nanocrystal model indicates a strong dependence of the band alignment type on the crystal size in accordance to the predictions in Reference \citenum{nanomain}. The detailed shape of small rutile nanocrystals also plays a crucial role. We showed that two types of rutile nanocrystal which both predominantly expose the (110) facets and which both have similar stability give rise to rather different behaviours in terms of band alignment. This fact also makes it difficult to establish robust rules of thumb when it comes to predicting band alignment in a rutile/anatase nanocrystal mixture and underlines the necessity for explicit simulations of those systems.

Using the efficiency of the DFTB method, we also investigated the effect of the anatase/rutile nanocrystal interface in the band alignment. We optimized the geometry of 156  anatase/rutile nanocrystal pairs where the mutual orientations were systematically varied. While, all interface models show the same type I alignment, the magnitude of the band-offset vary almost one electronvolt with different orientations. This observation suggests that it is important to consider interface effects in the band alignment and underpins the importance of being able to extend the reach of electronic structure simulations beyond the realms spanned by regular DFT.

\begin{acknowledgement}
The DFG grant RTG 2247 is acknowledged. JK would like to acknowledge the Swedish National Strategic e-Science programme (eSSENCE). We would also like to acknowledge Pavlin D. Mitev for his insightful comments. Prepared by LLNL under Contract DE-AC52-07NA27344. Project 20-SI-004 with Brandon Wood as PI was funded by the Laboratory Directed Research and Development Program at LLNL. 

\end{acknowledgement}





\bibliography{nanopaper}

\end{document}